# Spin wave damping
# in periodic and quasiperiodic magnonic structures

J. Rychły, J. W. Kłos and M. Krawczyk

Faculty of Physics, Adam Mickiewicz University in Poznań,
Umultowska 85, Poznań, 61-614, Poland

E-mail: rychly@amu.edu.pl; klos@amu.edu.pl

**Abstract.** We investigated the lifetime of spin wave eigenmodes in periodic and quasiperiodic sequences of Py and Co wires. Those materials differ significantly in damping coefficients, therefore, the spatial distribution of the mode's amplitude within the structure is important for the lifetime of collective spin wave excitations. Modes of the lower frequencies prefer to concentrate in Py wires, because of the lower FMR frequency for this material. This inhomogeneous distribution of amplitude of modes (with lower amplitude in material of higher damping and with higher amplitude in material of lower damping) is preferable for extending the lifetime of the collective excitations beyond the volume average of lifetimes for solid materials. We established the relation between the profile of the mode and its lifetime for periodic and quasiperiodic structures. We performed also the comparative studies in order to find the differences resulting from complexity of the structure and enhancement of localization in quasiperiodic system on the lifetime of spin waves.

1. Introduction

The one of the main factors limiting the application of spin waves (SWs) [1-3] in signal processing and transmission is damping, therefore, much effort is devoted to understanding the nature of SW damping[4-9] and to exploit the mechanisms of its compensation[10-12]. The microscopic models of damping are complex due to the variety of scattering processes at different time scales in which spin dynamics is involved. All of these effects (influencing directly or indirectly damping of the spin wave dynamics) can be included phenomenologically in Landau-Lifshitz equation (LLE) [13], if they act effectively in time scales in the range from fractions of ns to tens of ns (which correspond to the typical frequencies of spin wave precession). This classical equation describes damped precession of the magnetization in effective magnetic field. LLE is the equation for a continuous medium, where all atomic features of the system, related both to the structure and the physical mechanisms, are reflected in the continuous distribution of material parameters, describing different properties of the system. The internal components of the effective magnetic field: exchange, dipolar and anisotropy fields are expressed by material parameters such as: exchange length, saturation magnetization and anisotropy constant. The damping is included in the separate term of LLE, which is responsible for the torque, which drag the precessing magnetization towards the direction of the effective field. The magnitude of this torque and, as a result, the strength of damping, is set by the value of the empirical parameter called damping constant. This material parameter holds contributions from different energy transfer and dissipation processes on local scale.

In the magnetic inhomogeneous systems, like magnetic nanostructures composed of the elements made of different materials, damping parameter is spatially dependent. Definition of the effective damping parameter is not trivial for nano-structuralized materials. Damping of collective SW excitations in inhomogeneous system depends not only on its material composition but also on its geometry and spatial distribution of amplitude of SWs (which is non-uniform in nanostructures). Hence, the lifetime of SW excitation could be a better measure of the strength of damping for given eigenmode [14], than the damping parameter, because it can be directly obtained from the numerical solution in the frequency domain. The lifetime describes the characteristic time of the exponential decay of the amplitude of the mode. This parameter can be related to the linewidth in experimentally measured SW spectra [in ferromagnetic resonance (FMR), time resolved magneto-optical Kerr effect microscope (TR-MOKE) or Brillouin light scattering measurements (BLS)], therefore, we will use the lifetime to describe the influence of damping on collective SW dynamics in inhomogeneous magnetic nanostructures.

The experimental results for inhomogeneous (i.e., nanostructured) systems show the significant dependence of the linewidth on the mode profiles [15,16]. The lifetime depends on damping torque, which varies spatially both with the local value of damping parameter and with the amplitude of SW precession at the given position. The latter dependence can be explained by analyzing the damping term in the LLE, where the strength of damping torque increases with increase of the angle between the effective magnetic field and precessing magnetization vector. It is due to the presence of the cross product $\boldsymbol{M} \times \boldsymbol{H}_{eff}$. Thus the modes which will have the amplitudes concentrated in the parts of the systems, which are characterized by lower damping parameter, will be damped in smaller extend than the modes which are distributed more homogeneously. As a result, the effective lifetime (or damping parameter) cannot be calculated as a volume average of lifetimes (or damping parameters) of the constituent materials. This limitation is even valid for fundamental mode which also exhibits spatial deviation of the amplitude in different constituents due to different saturation magnetization, partial pinning and the presence of the demagnetizing fields.

In the presented paper we will establish the relation between the lifetime of SWs eigenmodes and their spatial distribution in magnonic structure, composed of two different ferromagnetic materials, differing also in damping. We will consider planar magnonic crystals (MCs) and quasicrystals (MQs) made of permalloy (Py) (low-damped material) and Co (high-damped material) wires [17-19]. Numerical calculations with finite element method (FEM) in the frequency domain will be performed for two sets of planar geometries differing in the ratio between in-plane dimension (width of wires, period) and out-of-plane dimension (thickness). The comparison between periodic and quasiperiodic systems will be presented to analyze the impact of the two effects, which enable to distinguish quasiperiodic from periodic system: presence of complex band spectrum and localization of bulk modes. We will discuss the change of lifetime within the magnonic bands and the global change of this parameter between different bands with the increase of frequency starting from the metamaterial limit, through the frequency regions, where the concentration in Py subsystems is observed, to higher frequency range, where the modes start to concentrate in Co wires.

The paper is organized as follows. After the introduction we will discuss: the geometry of the investigated systems, the assumed model, the numerical techniques and tools used to find the frequency spectrum and the profiles of modes. Then we will present and discuss the results – dependences of lifetime and concentration factor in Py on frequency. These dependences will be explained with the aid of plots presenting the profiles of modes. The publication will be summarized in separate section.

## 2. Model and structures

We considered planar structures composed of ferromagnetic wires arranged in periodic or Fibonacci (i.e., quasiperiodic) sequences – see Fig.1. The wires were made of Py and Co – metallic ferromagnets characterized by lower and higher damping, respectively. Both types of wires have the same dimensions. We investigated two regimes of sizes, both for periodic and quasiperiodic sequences, where the widths of wires (along the x-axis) were equal to 91 or 250 nm. The thickness of wires (along the y-axes) was the same for all considered by us systems (30 nm) and their length (along the z-axis) was assumed to be infinite. The wires were in the direct contact which ensured the exchange coupling between successive wires.

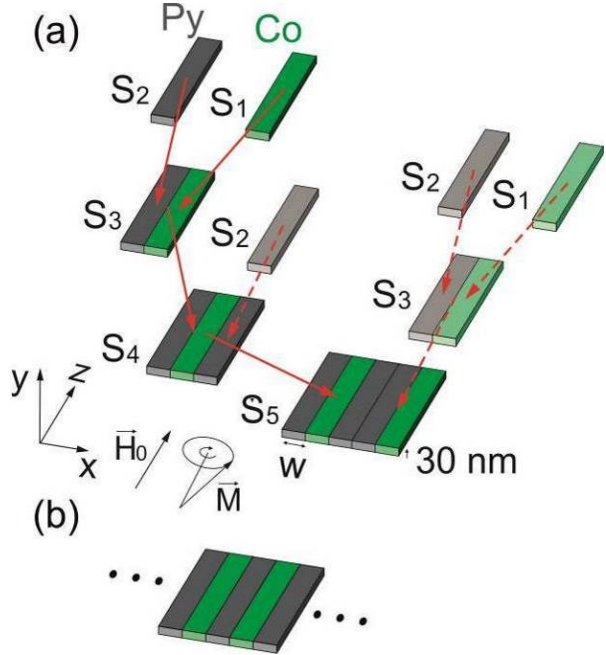

Figure 1. The considered magnonic systems: a) Fibonacci structure – magnonic quasicrystal (MQ) b) periodic structure – magnonic crystal (MC). The structures are planar and consist of infinitely long Py and Co wires being in the direct contact with each other. The external magnetic field $\mu_0 H_0=0.1$T is directed along the wires and is strong enough to saturate the samples. The section of MC in (b) has the same amount of wires as for the MQ presented in (a). We investigated the MC and MQ consisting of equal number of wires (i.e., 55) to compare properly the results obtained for both structures.

The quasiperiodic sequence of wires was achieved from the following Fibonacci recursion rule[20], that is: $S_{N+1}=[S_N+S_{N-1}]$, where the initial structures are: $S_1$ – single Co stripe and $S_2$ – single Py wire. Then, next structures, are constructed according to the mentioned rule by merging two previous structures (putting $S_N$ on the left and $S_{N-1}$ on the right to achieve the $S_{N+1}$ sequence). We followed this recursion procedure to obtain quasiperiodic (Fibonacci) sequence $S_{10}$, consisting of 55 wires (of two kinds: Co and Py). Because of the Fibonacci recursion rule, there will appear two Py wires next to each other within the structure. Although these wires are not separated and form one Py wire of double width, we will still count them as two Py wires.

To refer the results obtained for quasiperiodic structure to outcomes from periodic structure, we have investigated the section of periodic system composed of 55 Py and Co wires repeated alternately. Such number of wires is the same as in the considered $S_{10}$ Fibonacci sequence. We also assumed the same sizes of the wires (as for quasiperiodic structure) which results in the same total width of the sequence. The assumptions about sizes of periodic and quasiperiodic structures makes the strength of dipolar interactions comparable for both systems.

We assumed that the static magnetic configuration of considered structures is saturated by the external magnetic field applied in the direction of the wires axis. The field for which we have made our calculations was equal to $\mu_0 H_0 = 100$ mT. The material parameters of Co and Py were: saturation magnetizations $M_{Co} = 1.445 \times 10^6 \frac{A}{m}$, $M_{Py} = 0.86 \times 10^6 \frac{A}{m}$, exchange constants: $A_{Co} = 3.0 \times 10^{-11} \frac{J}{m}$, $A_{Py} = 1.3 \times 10^{-11} \frac{J}{m}$ and Landau damping parameter: $\alpha_{Co} = 0.1$ and $\alpha_{Py} = 0.01$, respectively. The gyromagnetic ratio $\gamma = 176 \frac{GHz}{T}$ was assumed the same for Co and Py wires.

To describe magnetization dynamics we have solved the LLE which is the equation of motion for the magnetization vector **M**:

$$\frac{d\mathbf{M}}{dt} = \gamma\mu_0 \left[\mathbf{M} \times \mathbf{H}_{eff} + \frac{\alpha}{M_S}\mathbf{M} \times (\mathbf{M} \times \mathbf{H}_{eff})\right], \tag{1}$$

where: $\alpha$ – is damping coefficient, $\mu_0$ – is permeability of vacuum, $\mathbf{H}_{eff}$ – is effective magnetic field. The first term in LLE describes precessional motion of the magnetization around direction of the effective magnetic field and the second term enrich that precession with damping. The effective magnetic field in general can consist of many terms, but in this paper we will consider only the most important components: external magnetic field $\mathbf{H_0}$, nonuniform exchange field $\mathbf{H}_{ex}$ and dipolar field $\mathbf{H}_d$: $\mathbf{H}_{eff} = \mathbf{H_0} + \mathbf{H}_{ex} + \mathbf{H}_d$.

We considered the linear regime of magnetization dynamics where we could clearly discuss the SW motion on the background of the static magnetic configuration in saturation state and investigate the SW eigenmodes in the system characterized by harmonic dynamics in time: $e^{i\omega t}$, where $\omega$ is the angular eigenfrequency. The LLE (1) for the magnetization vector **M** can be linearized in the form of set of two differential equations for complex amplitudes of dynamical components of magnetization $m_x$ and $m_y$:

$$(i\Omega - \alpha H_0)m_x(\mathbf{r}) =$$
$$= H_0 m_y(\mathbf{r}) - \frac{2}{\mu_0}\left(\alpha(\mathbf{r})\nabla \cdot \frac{A(\mathbf{r})}{M_S(\mathbf{r})}\nabla m_x(\mathbf{r}) + \nabla \cdot \frac{A(\mathbf{r})}{M_S(\mathbf{r})}\nabla m_y(\mathbf{r})\right) \tag{2}$$
$$+ M_S(\alpha(\mathbf{r})\frac{\partial}{\partial x} + \frac{\partial}{\partial y})\varphi(\mathbf{r}),$$

$$(i\Omega - \alpha H_0)m_y(\mathbf{r}) =$$
$$= -H_0 m_x(\mathbf{r}) + \frac{2}{\mu_0}\left(\nabla \cdot \frac{A(\mathbf{r})}{M_S(\mathbf{r})}\nabla m_x(\mathbf{r}) - \alpha(\mathbf{r})\nabla \cdot \frac{A(\mathbf{r})}{M_S(\mathbf{r})}\nabla m_y(\mathbf{r})\right) \tag{3}$$
$$- M_S(\frac{\partial}{\partial x} - \alpha(\mathbf{r})\frac{\partial}{\partial y})\varphi(\mathbf{r}),$$

where $\Omega = \frac{\omega}{\gamma\mu_0}$ is dimensionless frequency. The homogeneous external field has only z-component. $\mathbf{H_0} = [0,0,H_0]$. Exchange field for the system in saturation state has only nonzero dynamical components: $\mathbf{H}_{ex}(\mathbf{r},t) = [h_{ex,x}(\mathbf{r})e^{i\omega t}, h_{ex,y}(\mathbf{r})e^{i\omega t}, 0]$. The relation between the dynamical exchange field and the amplitudes of dynamical magnetization $\mathbf{m}(\mathbf{r},t) = \mathbf{m}(\mathbf{r})e^{i\omega t} =$ $[m_x(\mathbf{r})e^{i\omega t}, m_y(\mathbf{r})e^{i\omega t}, 0]$ can be express in linear approximation as [21]:

$$\mathbf{H}_{ex}(\mathbf{r},t) = \frac{2}{\mu_0 M_S(\mathbf{r})} \nabla \cdot \left(\frac{A(\mathbf{r})}{M_S(\mathbf{r})}\right) \nabla \mathbf{m}(\mathbf{r})e^{i\omega t}, \tag{4}$$

The second term on the right hand side of Eqs. (2) and (3) have exchange origin and results directly form the Eq.(4). Our calculations are based on FEM where the model of continuous medium is investigated. The exchange interaction between the successive stripes is included by appropriate boundary condition at the Co/Py interfaces. We used the natural boundary conditions resulting from the exchange operator (4) implemented in the linearized LLE equations (2-3). As was pointed out in [17, 22] these boundary conditions are: (i) continuity of dynamical components of magnetization and (ii) continuity of first derivatives of the dynamical components of magnetization multiplied by factors: $A/M_S$.

For considered geometry the static components of the demagnetizing field are equal to zero, nonzero are only $x$- and $y$-components of the dynamical dipolar field. Using the magnetostatic approximation, the demagnetizing field can be expressed as a gradient of the scalar magnetostatic potential:

$$\mathbf{H}_d(\mathbf{r},t) = [h_{d,x}(\mathbf{r})e^{i\omega t}, h_{d,y}(\mathbf{r})e^{i\omega t}, 0] = -\nabla\varphi(\mathbf{r})e^{i\omega t}. \tag{4}$$

With the aid of the Gauss equation, we obtained the following equation which relates magnetization and magnetostatic potential:

$$\nabla^2 \varphi(\mathbf{r}) - \frac{\partial m_x(\mathbf{r})}{\partial x} - \frac{\partial m_y(\mathbf{r})}{\partial y} = 0. \tag{5}$$

The Eqs. (4, 5) can be used to find dynamic components of demagnetizing field implemented already in equations (2-3), i.e., last terms on the right hand side of these equations.

The linearized LLE in the form of the eigenvalue problem (2-3) was solved by the use of FEM with the aid of COMSOL Multiphysics software [23, 24]. Unlike in our previous work[17], we do not neglect damping here and we studied influence of damping on SW spectra. From our calculations we have found the complex eigenfrequencies $f = \omega/2\pi = \text{Re}(f) + i\,\text{Im}(f)$, where the real part denotes the frequency of the SW and the imaginary part is connected to the lifetime $\tau$ of the SW modes by the relation: $\text{Im}(f)=(2\tau)^{-1}$[4,25]. The parameter $\tau$ describes the exponential decay (in time) of the energy carried be spin wave. Thus, from the numerical solutions (complex eigenfrequencies) we can get information about lifetime $\tau$ of each eigenmode.

To show how much of SW amplitude is concentrated in Py wires, we have introduced the measure of concentration for each mode, which we call concentration factor (CF) for Py. We defined this parameter according to the formula:

$$CF_\delta = \frac{\int_{Py\,stripes} |m_\delta|^2 ds}{\int_{Py+Co\,stripes} |m_\delta|^2 ds}, \tag{6}$$

where, in the numerator we have integrated the squared amplitude of the in-plane or out-of-plane component of the dynamical magnetization $m_\delta$ in the Py wires (where $\delta$ is $x$ or $y$, accordingly). The CF is normalized by the integral of the same kind calculated for the whole system. The symbol $ds$ in (6) denotes the element of area for $x$-$y$ cross-section of the structure. Thus the integrals in (6) are two-dimensional. The values of CF equal to 0 and 1 describe the mode which is concentrated entirely in Co and in Py, respectively. We will show in the next section that there is a strict correlation between the CF and the lifetime of the SWs in magnonic crystals and quasicrystals.

It is worth to notice, that in the considered range of frequencies the modes are almost homogeneous through the thickness of the structure [26] and the one-dimensional integration (i.e., along the *x*-direction) could be used in Eq.(6) as a good approximation. The modes start to be quantized through the thickness for higher frequencies than considered in this paper.

## 3. Results and discussion

We will discuss separately the outcomes for the systems composed of narrower ($w = 91$ nm) and wider ($w = 250$ nm) wires. The results for the structures with narrower wires are presented in Fig.2a-d where the lifetime and concentration factor for successive eigenmodes in dependence on frequency are showed. The gray and green points forming the horizontal lines in Fig.2a and b are the lifetimes for homogeneous films made of Py and Co, respectively. In this case, we performed calculations for uniform stripes of extended widths equal to the total width of the whole 55-elements sequence of wires. For the assumed direction of the external magnetic field (i.e., along the infinitely long wires) the modes quantized along the width of the structure are of the Damon-Eshbach type. For SWs of this type the spin wave lifetime $\tau$ in homogeneous film is independent on the frequency [25]. This peculiar feature is confirmed in our numerical calculations. This property is not inherited by nanostructures, where the dependence of the lifetime on frequency is observed. The red and dark blue points in Fig. 2a and b, representing lifetimes of the modes in nanopatterned films, are located between levels denoting the lifetimes of homogeneous films. They show evident dependence on the frequency with the two interesting trends: (i) global one – the lifetime is relatively short at lowest frequencies (in the metamaterial limit), then increases with increasing frequency, reaches maximum at about 18 GHz, and decay with the further increase of the frequency; (ii) local one – within each magnonic band (gaps between bands are marked by vertical gray stripes) the lifetime increases monotonically with frequency, reaching the highest (the lowest) value at the top (at the bottom) of the band. These trends are common both for periodic and quasiperiodic structures. There is one important difference between the frequency spectra of both structures. The spectrum consists of finite number of gaps and well defined bands for periodic system (Fig. 2b, d) and is determined by complex set of gaps for quasiperiodic system (Fig. 2a, c) [17]. Due to the complexity of the quasiperiodic structure we observe a fractal spectrum [17] with a lot of fine gaps and magnonic states between them. In considered periodic structure the first magnonic gap appears in the frequency range 16.5-19 GHz, while in MQ the relatively wide gap starts already at 14.5 GHz. For quasiperiodic system more modes have lifetime closer to the maximum value, than for a MC. We also noticed, that for the Fibonacci structure, the lowest modes have got slightly longer lifetimes than modes found in the periodic structure. It results from the different average composition of periodic and quasiperiodic structures build from Co and Py wires of the same dimensions. For periodic system the filling fraction of Py is equal to 50% (neglecting the ambiguities related to the selection of the first and the last wire in the sequence), while in the Fibonacci structure this parameter is more or less equal to golden number, which is approximately equal to 61.8%. This suggest that the lifetime for quasiperiodic system in metamaterial (long wavelength) regime should be larger and closer to the value characteristic for the solid Py film than for the periodic system. The lifetimes calculated numerically for homogeneous systems made of averaged material parameters: $A$, $M_S$ and $\alpha$ (with volume fraction of Py: 50% like in the periodic structure and 61.8% like in the quasiperiodic one) confirm this suggestion, but they are noticeably overestimated (0.97 ns and 1.24 ns for smaller and larger filling fraction, respectively) in relation to the lifetimes of the lowest (i.e., fundamental) mode in magnonic crystal and quasicrystal, respectively (Fig. 2a and b). Thus, there is need to investigate this structurally induced change of damping in more systematic way with detailed analyses of the distribution of SW amplitude.

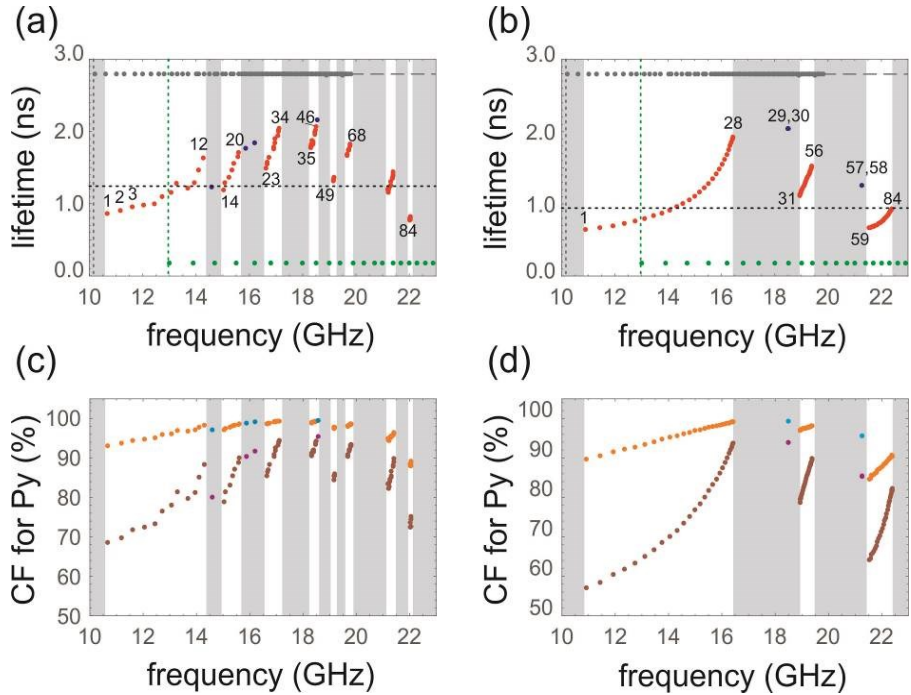

Figure 2. (a,b) Lifetime of SW eigenmodes versus their frequencies for (a) Fibonacci structure - MQ and (b) periodic structure – MC, consisting of 55 wires of 91 nm width. (c, d) The concentration factor (CF) for $m_x$ component (brown and violet points) and $m_y$ component (orange and light blue points) of SW eigenmodes, showing the strength of the excitation within the Py subsystem for (c) MQ and (d) MC. The grey areas mark the most pronounced frequency gaps. Green (grey) horizontal series of points show the lifetime of eigenmodes obtained for solid Co (Py) homogeneous film of the same thickness and external width as considered for MC and MQ structures. The dashed vertical green and grey lines mark the frequency of the first FMR mode obtained for homogeneous Co and Py film, respectively. Dark blue, light blue and violet points which can be found in the gaps show the lifetime and CF of selected surface modes. Red, orange and brown points show the lifetime and CF of the rest of the modes (mostly bulk modes) found for Fibonacci and periodic structure. The dashed horizontal lines mark the lifetime of the homogeneous stripe of averaged material parameters.

To relate the lifetime of the modes with the spatial distribution of their profiles, we will plot (in Fig.2c,d) the concentration factors in Py system: $CF_x$ and $CF_y$, defined by Eq. (6) for $m_x$ and $m_y$ components of dynamical magnetization, respectively. The dependences of $CF_x$ (brown points) and $CF_y$ (orange points) on frequency are qualitatively the same, but differ in the values they reach – the $CF_y$ is noticeable bigger than $CF_x$. The difference between concentration factors $CF_x$ and $CF_y$ results from the symmetry of the system. Due to differences in $x$- and $y$-components of dynamical dipolar fields, the component $m_y$ is strongly reduced in reference to $m_x$ component (ellipticity of precession is larger) in Co stripes, than in Py stripes.

We expect that modes of higher concentration in Py will be characterized by longer lifetime than those concentrated in Co. For these modes the damping torque is low both in Py and Co subsystem. Small damping torque in Py appear because of the low value of intrinsic damping (small $\alpha$), while the reduced damping torque in Co results from the small amplitude of the SW precession (and low value of $\boldsymbol{M} \times \boldsymbol{H}_{\text{eff}}$ – see Eq. (1)). Fig.2.c and 2.d presents the dependence of CF on frequency for periodic and quasiperiodic systems, respectively. Both, the global change of the lifetime in the function of frequency and the local changes of this parameter within each band, are clearly reflected in the dependence of the CF on frequency. This supports our explanation of non-local reduction of damping in Co/Py nanostructures, resulting from different distribution of the amplitude of SWs between Co and Py subsystems. It is worth to notice that the relation between CF and lifetime is not linear, which results from nonlinear relation between precession amplitude and damping torque. CF can also reflect some other peculiarities of mode profiles. All these effects can result in the lack of one-to-one correspondence between the CF and the lifetime. For instance, the ranges of $CF_x$ for the second and the third magnonic bands, overlap (see Fig. 2d) but their ranges of corresponding lifetimes are separated – it means that we can have two states of the same $CF_x$, which differ in the lifetime. The other problem is some ambiguity in the definition of CF, which is related to the measure used in Eq. (2) – we choose the most intuitive formulation of CF – we have integrated the squared absolute value of $|m_x|^2$ (and not e.g., the absolute amplitude $|m_x|$). This particular choice affect only quantitatively on the dependence of CF on frequency and can be justified by the relation of experimental characteristics to the squared amplitude of SWs. Overall, the calculated CF explains the general change of lifetime in dependence on frequency.

For further analysis of the impact of the shape of profiles on the lifetime of SWs, we plotted the selected modes from quasiperiodic (Fig.3) and periodic sequences (Fig.4). We pay attention on the location of the nodes and on the position of the areas of the SWs amplitude concentration. We start our discussion from the description of the lowest modes in the frequency scale i.e., in the metamaterial regime. In this frequency range the modes vary in the scale of the whole structure, being relatively smooth on the distances comparable to the sizes of the single elements composing the nanostructure. Thus, the lowest SW modes have very little oscillations and nodes. Their amplitude (both in Py and Co subsystem) goes rarely to zero and must follow the long scale oscillations characteristic for metamaterial regime (see modes 1, 2, 3 in Figs. 3 and 4). As a result of relatively high amplitude of SWs also in Co, the lifetimes of the lowest modes are short.

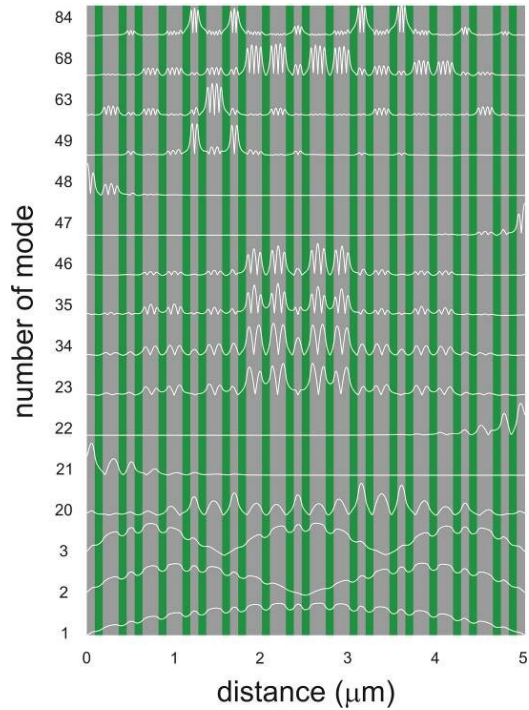

Figure 3. The profiles of selected modes for Fibonacci structure consisting of 55 wires of the 91 nm width (see Fig. 2a). We plotted the amplitudes of the in-plane component of the dynamical component of magnetization $|m_x|$ versus distance along the whole width of the structure. The modes are presented on the background, where green and gray colors denote the areas of Co and Py wires, respectively.

When we look into the details of the profiles of the lowest modes, we see that the amplitude in Co is slightly lowered in comparison to the amplitude in Py. However, this finding do not undermine the previous conclusion about high damping of the lowest modes. The lowering of the SW amplitude in Co can be easily understood when we notice that the frequencies of these modes are below the FMR frequency of Co. It means that the SW amplitude will decay into the center of Co wires from the forced oscillations at the Py edges (only exponential decay is accepted below FMR frequency). This slight lowering of the amplitude in Co subsystem could suggest that the lowest modes will have longer lifetime in comparison to the modes in homogeneous system (formed by averaging of the material parameters, including the damping constant), where the amplitude changes smoothly. But, in fact, the opposite is true: the lifetime of fundamental mode for nanostructure is slightly lower than for the average system. This finding demonstrate, that the volume averaging of the damping constant is not fully justified approach in magnetic composites, even in the long wavelength limit. The lifetimes of higher modes, starting from the fundamental excitation, grow with frequency because of the increase of number of nodes which are usually located in Co. This reduce the damping torque in Co and extends the lifetime (see e.g., mode 20 in Fig. 3 and mode 28 in Fig. 4).

We will focus now on higher frequency SWs (from higher bands) in the MC. In the second band (modes 31-56 in Fig.4) all modes have one node in every Py wire. This reduces the averaged amplitude in Py subsystem as compared to the excitations of the first band. As a result of increasing the contribution of SWs excited in Co, the lifetime for SWs in the 2nd band will be shortened. The similar arguments can be used to explain the reduction of the lifetime in the third band (where two nodal lines in each Py wire are observed) in reference to the second one.

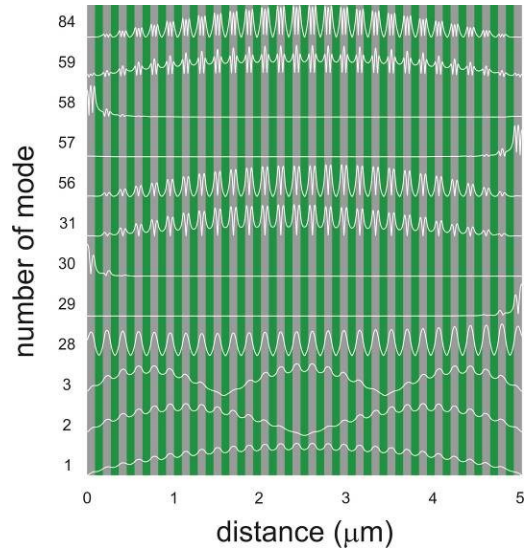

Figure 4. The profiles of selected modes for periodic structure consisting of 55 wires of the 91 nm width (see Fig. 2b). We plotted the amplitudes of the in-plane component of the dynamical magnetization, $|m_x|$ versus distance along the whole width of the structure. Modes are presented on the background, where green and gray colors denote the areas of Co and Py wires, respectively.

The scenario for changes of lifetime within the second band is similar to the first band: while in all Py wires one node is observed, then the successive modes have more and more Co wires with nodes; this gradually increases the lifetime. We can explain similarly the changes of modes profiles in the third band (modes 59-84), where two nodal lines in each Py wire appear. This will led us to the same conclusion about the changes of lifetime within the third magnonic band as for the previous ones.

Presented mechanism seems to be systematic and can be used to explain the increase of the lifetime of the modes within the bands with increasing frequency and reduction of the lifetime of higher band in reference to lower one. However, this clear picture is not valid for high frequency bands, where SWs start to oscillate spatially in Co. The harmonic oscillations in Co stripes appear at higher frequencies, well above FMR frequency for Co, due to the finite width of the wires. Changes of the spatial distribution of SWs amplitudes are much more complex with increasing frequency and are difficult to describe in the frame of simple mechanism. We will investigate them briefly for the wires of larger widths in the last part of this section.

We will now discuss the relation between SWs mode profiles above the metamaterial regime frequencies in the quasiperiodic structure (i.e., above third mode in Fig.3). There are two important structural differences in the reference to a periodic sequence that are worth to stress. The first difference is that in the considered quasiperiodic structures there are Py wires both of single and of double width. The second difference is lack of periodic repetition of elements in the structure. Due to two different widths of wires the modes can concentrate in both kinds of Py wires. The lower modes (see e.g., the modes: 23, 34, 35 and 46 in Fig. 3) are concentrated mostly in wider Py wires, higher modes start to be excited in the single Py wires (e.g. modes 49 and 84). Thus, the quantization of SWs within Py wires depends also on the width of wires. In single Py wires the increase of the number of nodes (or spatial oscillations) appears with larger differences of frequencies – e.g., the group of modes with one node (e.g., the mode no. 49 shown in Fig. 3) and the group with two nodes in Py wires (e.g., the mode 84) are more separated in the frequency domain than the group of modes concentrated in double wires (in double Py wires) – e.g., the mode 23 in Fig. 3 originating from the group with one node in double Py wire and mode 35 with two nodal points in double Py wire.

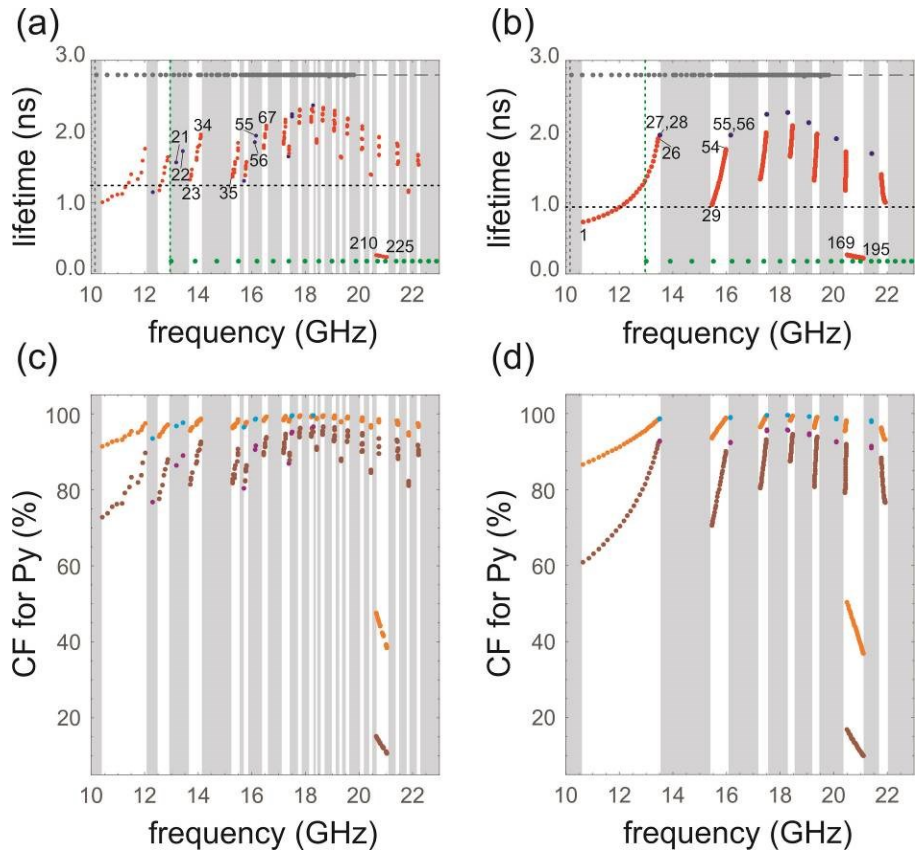

Figure 5. (a,b) Lifetime of spin wave eigenmodes versus their frequencies for (a) Fibonacci structure - MQ and (b) periodic structure – MC, consisting of 55 wires of 250nm width. (c,d) The concentration factor (CF) for $m_x$ component (brown and violet points) and $m_y$ component (orange and light blue points) of spin wave eigenmodes showing the strength of the excitation within the Py subsystem for (c) MQ and (d) MC. The grey areas mark the most pronounced frequency gaps. Green (grey) horizontal series of points show the lifetime of eigenmodes obtained for solid Co (Py) layer of the same external thickness and width as considered for MC and MQ structures. The dashed vertical green and grey lines mark the frequency of the first FMR mode obtained for homogeneous Co and Py stripe, respectively. Dark blue, light blue and violet points, which can be found in the gaps, show the lifetime and CF of selected surface modes. Red, orange and brown points show the lifetime and CF of the rest of the modes (mostly bulk modes) found for Fibonacci and periodic structure. The dashed horizontal lines mark the lifetime of the homogeneous stripe of averaged material parameters.

The lack of regular arrangement of elements (sets of wires) in quasiperiodic structure means that every Py stripe can have unique surrounding [17]. This strengths the localization of SWs, due to very few (if any) equivalent locations for SW eigenmodes (see modes with numbers larger than 23 in Fig. 3). The small number of Py wires with significant amplitude of modes, protects the distribution of the modes across the whole structure. It is worth to notice that eigenmode in quasiperiodic structure can be formed by the excitation in a few Py wires only if they have similar surrounding of the other stripes [17].

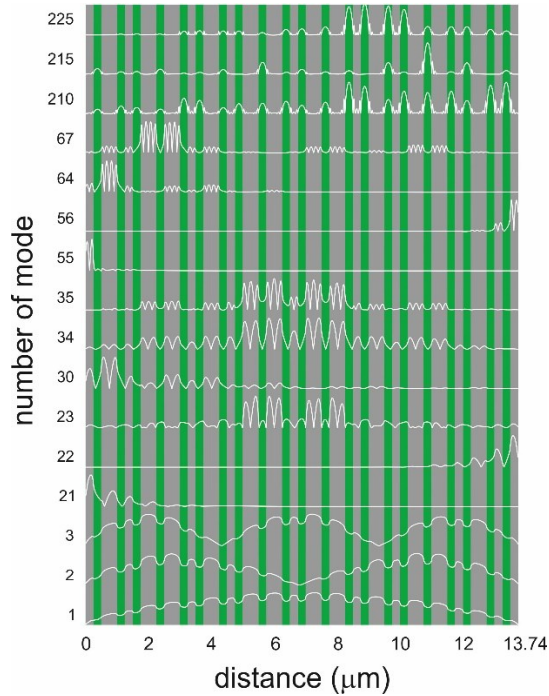

Figure 6. The profiles of the selected modes for Fibonacci structure consisting of 55 wires of 250 nm width (see Fig. 5a). We plotted the amplitudes of the in-plane component of the dynamical magnetization, $m_x$, versus distance along the whole width of the structure. Modes are presented on the background, where green and gray colors denote the areas of Co and Py wires, respectively.

For the Fibonacci quasiperiodic structure it is not easy to analyze the frequency spectrum, because it is impossible to point out the allowed frequency bands. We can only find its counterpart by the complement of the complex, fractal set of magnonic gaps [17]. For the finite structure the magnonic gaps can be found as a frequency ranges of the width much bigger than average frequency interval between successive modes. However, this limits the detection of the fine magnonic gaps, therefore, in Fig. 2a and c we identified only the widest magnonic gaps in the spectrum.

In the case of MQs the systematic discussion of the changes of mode profiles and their lifetimes is difficult to do in the manner presented for the MC. We selected for discussion only some modes at the edges of the largest magnonic gaps (modes: 20, 23, 34, 35, 46, 49, 63, 68 and 84 in Fig. 3). These modes differ significantly in the spatial distribution of SW amplitude which results in noticeable differences in their lifetime. We found three main features of the profiles of modes in quasiperiodic system which affect clearly the lifetime of those modes. These are: (i) the ratio of the number of double to single Py wires in which high amplitude of the SW mode is concentrated, (ii) the number of nodes in Co wires (especially in those located next to the Py wires, in which the amplitude concentrates), (iii) the number of nodes in Py wires. The best composition of the mentioned conditions, which will give us the guidance to pick up the profile of the mode characterized by the long lifetime, is: the mode should be concentrated in the double Py wires with low number of nodes inside of these wires and should decay rapidly in the surrounding Co wires, which can be ensured by the nodes located in Co. If we compare e.g., the modes 34 and 84, we will be able to explain the differences in the lifetimes of those modes (short lifetime for mode 84 and long lifetime for mode 34) in the terms of the discussed above features.

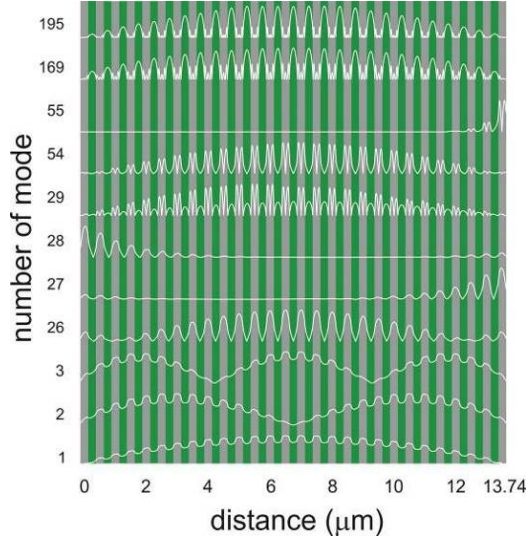

Figure 7. The profiles of selected modes for periodic structure consisting of 55 wires of the 250nm width (see Fig.5b). We plotted the amplitudes of the in-plane $m_x$ component of the dynamical magnetization versus distance along the whole width of the structure. Modes are presented on the background, where green and gray colors denote the areas of Co and Py wires, respectively.

The considered systems are finite sequences of elements (wires) ordered in periodic or quasiperiodic manner. For such structures both the bulk states (forming the frequency bands) and surface states (appearing in frequency gaps) can exist. The frequency gaps are the frequency ranges forbidden for the bulk modes. However surface states, which are localized on the external wires (and exponentially decaying inside of the structure), can exist in these frequency ranges. For the structures under consideration, the external stripes are made of Py (both for periodic and quasiperiodic structures), therefore, the most of the amplitude of the surface modes is concentrated in the material of low damping (Py). This explains the relatively long lifetime of these modes (Fig. 2a and b). It is worth to note that, due to the significant width of the structure (and big separation between surfaces), two surface modes appearing in one gap are practically degenerate and their lifetimes are practically the same, although in the quasicrystal the y-axis of symmetry is not present.

We are going to discuss now the results obtained for the larger structures i.e., for the systems made of wider wires. The dependences of the lifetime and CF for the structures build from wires of 250 nm width were set together in Fig. 5. We chose to use the style of points, lines and the ranges of frequency and lifetime the same as in Fig. 2 to make the comparison of the plots presented in Fig. 2 and Fig. 5 easier. Comparing obtained results for the structures made of narrower wires (Fig. 2) to the results for wider wires (Fig. 5), is easy to notice that the lifetime of eigenmodes changes in similar limits in the same frequency ranges. We can make the same observation for dependences of CF on frequency, which is also presented in Fig. 2 and Fig. 5.

The one most distinctive difference between the spectra for the systems presented in Fig. 2 and Fig. 5 is a difference in the number of magnonic gaps and bands. It is because of the differences of the widths of wires. For the periodic system the width of wires (and as a result the period) determines the size of the first Brillouin zone and location of frequency gaps. The system of wider wires is characterized by finer frequency spectrum. This rule can be also generalized for quasiperiodic systems. In our study we keep the total number of wires within the structure constant (i.e., equal to 55), therefore, the number of modes per one band is the same for the structures with narrower and wider wires.

The general trend of the dependence of lifetime and CF versus frequency is more or less the same for each of considered structures: the envelope goes at first up, reaches maximum at around 18 GHz and then it starts to fell down. This dependence is more dynamic (i.e., has more steep slopes) for structures build from wires of 250 nm width. In the considered range of frequencies we obtain for periodic sequence of narrower wires (Fig. 2b, d) 3 bands and 3 gaps, for periodic sequence of wider wires (Fig. 5b, d) – 8 bands and 8 gaps. For quasiperiodic systems we marked only the widest gaps but the increase of the number (or density in frequency domain) of magnonic gaps is evident for the structure with wider wires (Fig. 5a, c). We noticed that the modes in the structures, constructed from wider wires, are on average less damped (have got slightly longer lifetime) than the modes for systems build with narrower wires.

The profiles of modes for structures with wider wires are presented in Fig. 6 (for quasiperiodic structure) and in Fig. 7 (for periodic structure). In general, the profiles of modes for periodic and quasiperiodic systems look quite similar to those presented in Figs. 3 and 4. The lowest modes (i.e., the modes in the metamaterial regime) look practically the same as their counterparts in Fig. 3 and Fig. 4. Some differences in the shape of the corresponding modes from the structures differing in the width of wires can be attributed to different frequencies and different relation between dipolar and exchange interactions – compare the last modes in the first band for periodic sequences of narrower wires (mode 28 in Fig.4) and wider wires (mode 26 in Fig.7).

The important difference between Figs. 2 and 5 for the systems with narrow and wide wires, respectively, is the presence, in the later system, of the band of modes which are almost entirely concentrated in the Co wires. For periodic (quasiperiodic) sequence we observe the band of modes no. 169-195 (210-225) which are highly damped and characterized by the lifetimes close to the lifetime of homogeneous film of Co – see Fig.5a,b. The selected profiles are presented in Fig. 6 and Fig. 7.

For each of the considered structures similar dependence of the lifetime and CF on frequency within the magnonic bands can be found: the modes at the bottom of the band are the most damped and then their damping is gradually reduced with the increase of the frequency. The only exception from this rule are the bands of modes concentrated in Co, where the lifetime of modes is reduced with increasing frequency.

All considered structures are terminated with Py stripes. Due to the low damping of Py, the surface modes (mostly concentrated in outermost Py stripes) are characterized by relatively long lifetimes.

We also inspected carefully the bottoms of magnonic bands to check how SWs penetrate Co wires, which is the main source of the increase of damping for modes concentrated mostly in Py. We found that for some modes (e.g., mode 29 for periodic sequence in Fig.7) the amplitude of SWs do not decay in the center of Co wires and, therefore, these modes can be considered as modes supporting the SW dynamics both in Py and Co subsystem.

## 4. Conclusions

We have showed that SW lifetimes in one-dimensional planar MCs and MQs systems differ substantially from damped SWs in planar homogeneous systems. We have investigated nanostructures composed of the two materials: the one characterized by lower saturation magnetization and lower damping and the other one with higher saturation magnetization and higher damping. For such systems the lifetime of eigenmodes can be extended over the lifetime of modes in homogeneous slab, made of artificial material, characterized by parameters, being the volume average of parameters for Co and Py. The frequency dependence of lifetime for nanostructure is not constant, as it is for a homogeneous layer in a Damon-Eshbach configuration [25], but shows specific dependence, having the common features for periodic and quasiperiodic bandgap materials, and for the two different aspect ratios of the xy-cross-section of Co and Py wires. For the lowest modes the lifetime is slightly below the value of this parameter for homogeneous layer of averaged material parameters, then increases, and after reaching the maximum, decays gradually. These changes take place in the frequency range, corresponding to the lowest magnonic bands, where the modes are concentrated mostly in the low damped material – Py. For higher frequencies we start to observe (for the structures with 250 nm wide wires) the bands of the modes which are concentrated in Co and characterized by very short lifetime. Those bands are out of the described above trend.

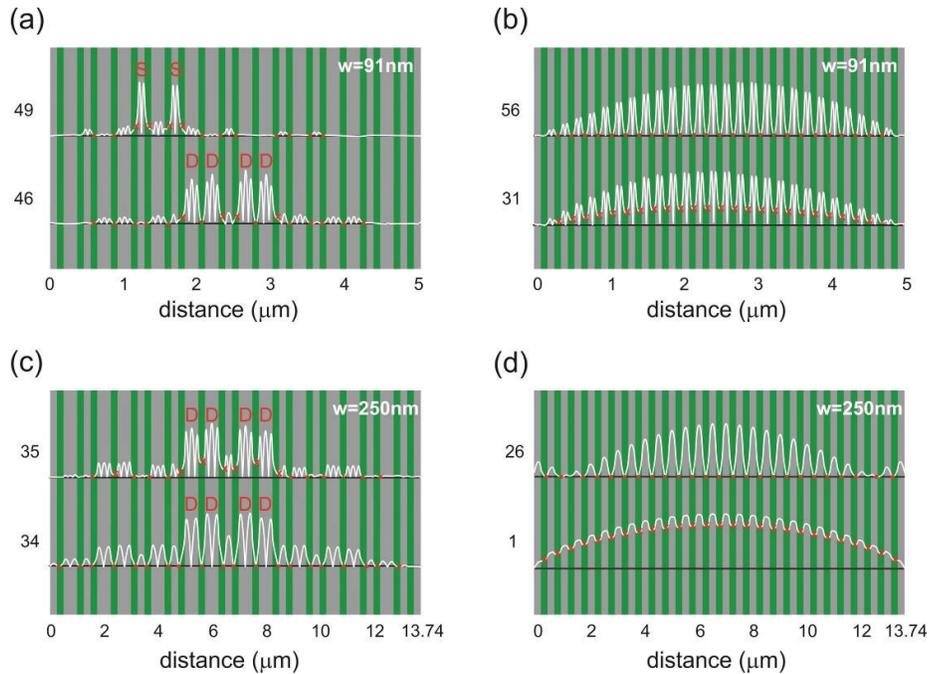

Figure 8. The chosen modes for the Fibonacci (a,c) and periodic (b,d) structures constructed from wires of two different widths: 91nm (a,b) and 250nm (c,d). The distance shows the width of the investigated structures. The number of selected modes are placed on the left side of each subfigure. By the red circles are marked the nodes in cobalt wires; the red crosses denotes the cobalt wires in which the amplitude of mode does not fell down to zero (has got nonzero value). The capital 'D' and 'S' letters mark the double and the single wires of Py in Fibonacci structures.

Our main finding is the description of the relation between spatial distribution of amplitude of the SW eigenmodes and their lifetimes. We were able to correlate the concentration of the mode in Py (or Co) subsystem (expressed by introduced measure of this concentration) to its lifetime. We also found some specific features of the profiles of modes which are useful in qualitative determination of the lifetime of particular mode. We summarized this analysis in Fig.8. In the low frequency range, where the modes are concentrated mostly in Py, the one important factor is the number of nodes of the profile located in Co wires. For instance, the modes 56, 34, 26, presented in Fig.8b,c,d, respectively, are characterized by the longer lifetime due to reduction of SW precession in Co, manifested by large number of nodes located in Co. For quasiperiodic sequence, the type of Py wires (single or double ones) in which the SW is concentrated, plays important role. Modes concentrated in the double Py wires (e.g., mode 46 in Fig. 8a) have got longer lifetimes than those concentrated in the single Py wires (e.g., mode 49 in Fig.8a). This relation between the spatial profiles of SW and its lifetime allowed us to explain the main trend in dependence of lifetime on frequency and to deduce some detailed features of this dependence: the monotonic growth of the lifetime within individual band (for periodic systems) or the jumps of lifetime across the gaps (for periodic and quasiperiodic systems).

We showed that the damping in magnonic nanostructures can be molded by introducing the periodic as well as quasiperiodic modulation. The periodic and quasiperiodic geometry induces additional dependence of damping on frequency absent in homogeneous planar systems, which can be tuned by the change of the structural parameters. This property opens the possibility to design magnonic devices [27, 28, 29, 30], for which the reduced damping will be gained for specific spin wave modes or in selected frequency ranges.

## 5. Acknowledgements

The research has received funding from Polish National Science Centre project DEC-2 12/07/E/ST3/00538, and from the EUs Horizon2020 research and innovation programme under the Marie Sklodowska-Curie GA No644348 (MagIC). We would like to thank prof. V. N. Krivoruchko for fruitful discussion and valuable suggestions.

## 6. Literature

[1] Krawczyk M and Grundler D 2014 *J. Phys.: Condens. Matter* **26** 123202
[2] Kruglyak V V, Demokritov S O, and Grundler D 2010 *J. Phys. D: Appl. Phys.* **43** 264001
[3] Demokritov S O and Slavin A N (eds.) 2013 *Magnonics from fundamentals to applications* (Berlin: Springer)
[4] Gurevich A G and Melkov D A 1996 *Magnetization Oscillations and Waves* (CRC Press)
[5] Garanin D A 1997 *Phys. Rev. B* **55**, 3050
[6] Tserkovnyak Y, Brataas A, Bauer G E W, and Halperin B I 2005 *Rev. Mod. Phys.* **77**, 1375
[7] Steiauf D and M. Fähnle M 2005 *Phys. Rev. B* **72**, 064450
[8] Krivoruchko V N 2015 *Low. Temp. Phys.* **41** 864
[9] Wang W, Dvornik M, Bisotti M-A, Chernyshenko D, Beg M, Albert M, Vansteenkiste A, Waeyenberge B V, Kuchko A N, Kruglyak V V, and Fangohr H 2015 *Phys. Rev. B* **92**, 054430

[10] Ando K, Takahashi S, Harii K, Sasage K, Ieda J, Maekawa S, and Saitoh E 2008 *Phys. Rev. Lett.* **101**, 036601
[11] Zhou Y, Jiao H J, Chen Y, Bauer G E W, and Xiao J 2013 *Phys. Rev. B* **88**, 184403
[12] Baláž P and Barnaś J 2015 *Phys. Rev. B* **91**, 104415
[13] Landau L and Lifshitz E 1935 *Phys. Z. Sovietunion* **8**, 153
[14] Romero Vivas J, Mamica S, Krawczyk M and Kruglyak VV 2012 *Phys. Rev. B* **86**, 144417
[15] Shaw J M, Silva T J, Schneider M L, and McMichael R D 2009 *Phys. Rev. B* **79**, 184404
[16] Nembach H T, Shaw J M, Boone C T, and Silva T J 2013 *Phys. Rev. Lett.* **110**, 117201
[17] Rychły J, Kłos J W, Mruczkiewicz M, and Krawczyk M 2015 *Phys. Rev. B* **92**, 054414
[18] Wang Z K, Zhang V L, Lim H S, Ng S C, Kuok M H, Jain S, and Adeyeye A O 2009 *Appl. Phys. Lett.* **94** 083112
[19] Sokolovskyy M L and Krawczyk M 2011 *J. Nanopart. Res.* **13** 6085
[20] Janot C, *Quasicrystals: a Primer, 2nd ed*. 1994 (Oxford University Press)
[21] Krawczyk M, Sokolovskyy M L, Klos J W and Mamica S 2012 *Adv. Cond. Matt. Phys.* **2012**, 1 (doi: 10.1155/2012/764783)
[22] Wang Z K, Zhang V L, Lim H S, Ng S C, Kuok M H, Jain S and Adeyeye A O 2010 ACS Nano **4**, 643
[23] Mruczkiewicz M, Krawczyk M, Sakharov V K, Khivintsev Yu V, Filimonov Yu A, and Nikitov S A 2013 *J. Appl. Phys.* **113** 093908
[24] Mruczkiewicz M, Krawczyk M, Gubbiotti G, Tacchi S, Filimonov Yu A, Kalyabin D V, Lisenkov I V and Nikitov S A 2013 *New J. Phys.* **15** 113023
[25] Stancil D, Prabhakar A 2009 *Spin waves – theory and applications* (Springer)
[26] Rychły J, Gruszecki P, Mruczkiewicz M, Kłos J W, Mamica S, and Krawczyk M 2015 *Low. Temp. Phys.* **41** 959
[27] Nikitin A A, et al. 2015 *Appl. Phys. Lett.* **106** 102405
[28] Chumak A V, Serga A A and Hillebrands B 2014 *Nat. Commun.* **5** 4700
[29] Inoue M, Baryshev A, Takagi H, Lim P B, Hatafuku K, Noda J and Togo K 2011 *Appl. Phys. Lett.* **98** 132511
[30] Metaxas P J, et al. 2015 *Appl. Phys. Lett.* **106** 232406